\documentclass[aps,prl,twocolumn,showpacs]{revtex4}
\usepackage{bm}
\usepackage{amsmath}
\usepackage{amssymb}
\usepackage{graphicx}

\def\bk{{\bf k}}

\def\bl{{\bf l}}

\begin{document}

\title{dc Conductivity of an array of Josephson junctions in the insulating state}
\author{S.~V. Syzranov$^1$, K.~B. Efetov$^1$, and B.~L. Altshuler$^{2}$}
\affiliation{$^1$Theoretische Physik III, Ruhr-Universit\"at Bochum, D-44801 Bochum,
Germany\\
$^2$Physics Department, Columbia University, New York, N.Y. 10027, USA}
\date{\today}

\begin{abstract}
We consider microscopically low-temperature transport in weakly
disordered arrays of Josephson junctions in the Coulomb blockade
regime. We demonstrate that at sufficiently low temperatures the
main contribution to the dc conductivity comes from the motion of
single-Cooper-pair excitations, scattered by irregularities in the
array. Being proportional to the concentration of the excitations,
the conductivity is exponentially small in temperature with the
activation energy close to the charging energy of a Cooper pair on a
superconductive island. Applying a diagrammatic approach to treat
the disorder potential we calculate the Drude-like conductivity and
obtain weak localization corrections. At sufficiently low
temperatures or strong disorder the Anderson localization of Cooper
pairs ensues.
\end{abstract}

 \maketitle

Artificially fabricated Josephson junction arrays (JJA) reveal various
fundamental quantum phenomena ranging from quantum phase transitions to the
motion of single charges and vortices~(see, e.g. \cite{Fazio:SITReview}).
The flexibility of their design and parameters makes JJAs a perfect
laboratory for study of physics underlying these phenomena.

Low-temperature transport in the array is determined by the ratio of
the characteristic Josephson coupling energy $J$ to the effective
charging energy $B/2$ of adding one Cooper pair on a superconductive
island in the array. In the two limiting cases, $J> J_{c}$ and
$J<J_{c},$ where $J_{c}\sim $ $B$, the JJA is known to be
macroscopically superconducting or insulating, respectively, which
has been demonstrated by a microscopic calculation quite long ago
\cite{Efetov:fundamental}. Especially interesting is the
two-dimensional ($2D$) case when a Josephson- or charging-energy-
dominated array may undergo respectively a vortex- or charge-
unbinding Berezinskii-Kosterlitz-Thouless (BKT) \cite{BKT}
transition to a normal conducting state.

Extensive studies of JJA dynamics have been carried out since the
first array fabrication~\citep{Voss:firstarray}. However\textbf{,}
transport properties of the Coulomb-blockaded JJAs are not fully
understood yet, although the issue of the conductivity of a
granulated superconductor in the insulating phase has first been
addressed already in Ref.~[\onlinecite{Efetov:fundamental}]. It was
demonstrated that the frequency-dependent conductivity $\sigma
\left( \omega \right) $ {had} sharp peaks at frequencies
corresponding to the excitation energies of Cooper pairs. However,
the \textit{dc}
 conductivity in the insulating state has not been calculated
explicitly.

Later, the conductivity of large JJAs has been studied close to the
superconductor-insulator transition using mean-field-type
approaches~\cite{Bruder:meanfieldconductivity} or scaling arguments
based on the charge-vortex duality \cite{Fisher:unicondmagnetic}. In
many respects the study of transport was phenomenological and,
particularly, did not account properly for the {effects} of
disorder.

{
 However, as the relaxation of charge
carrying excitations due to inelastic processes strongly decreases
with temperature, the disorder must play the major role in the
low-temperature transport in the insulating phase of the array. }


Actually, disorder is intrinsically present in the conventional
tunnelling Hamiltonian describing the coupling between
superconducting islands, as electrons can hop from one island to any
state near the Fermi surface in another island. Such a disorder
determines the conductivity of a regular array of normal metal
grains (see, e.g., Ref.~[\onlinecite{Beloborodov:grainreview}]). Is
the same true also for the insulating state of a regular array of
superconducting islands or the macroscopic irregularities in the
system should be accounted for? How does the conductivity depend on
the temperature and on the range of electron-electron interactions?

In the present Letter we study a large JJA deeply in the insulating
state ($J\ll B/2$) and address these questions. Most of our results
are valid for any dimensionality of the array, although at some
points we restrict ourselves to $2D$. We assume for simplicity that
the superconducting gap $\Delta $ in a single island is the largest
energy scale.

We calculate the conductivity of the JJA under rather general
assumptions and demonstrate that the conductivity of a regular
periodic array without a macroscopic disorder remains infinitely
large as long as macroscopic disorder and inelastic processes are
neglected. The \textit{dc} is carried mainly by single-Cooper-pair
excitations with the charge $\pm 2e$. Such bosonic particles move in
a regular array without being scattered. At the same time, the
density of the bosons is exponentially low in temperature, $\propto
\exp (-E_{0}/T)$, with $E_{0}$ close to $B/2$.

Macroscopic disorder in the JJA results in the boson scattering on
irregularities and makes the conductivity finite. The conductivity
is proportional to the density of the bosons, i.e to
$\exp(-E_{0}/T)$, in the limit $T\ll E_{d}$
\begin{equation}
\sigma \sim e^{2}T^{-1}\tau (\min (T,J))^{2}\exp (-{E_{0}}/{T}),
\label{a1}
\end{equation}%
where $\tau $ is the elastic scattering time and $E_{d}$ is the
energy of two bosons of opposite charge (boson dipole) located on
neighboring islands. The energy $E_{d}$ is either of the same order
as $E_{0}$ or considerably smaller depending on the range of the
effective Coulomb interaction between the bosons, determined by the
capacitive properties of the array. Eq.~(\ref{a1}) is an analogue of
the classical expression for the conductivity of free particles. In
the limit $T\ll J$ the {pre-exponential} can be evaluated exactly
and equals $4e^{2}T\tau /\pi $.

Remarkably, the description in terms of scattered bosons allows one not only
to obtain the classical limit, Eq.~(\ref{a1}), but also to describe the
quantum interference leading to localization effects. In $2D$ the first weak
localization correction $\delta \sigma _{WL}$ to $\sigma $ takes the form
\begin{equation}
\delta \sigma _{WL}\sim -e^{2}T^{-1}\min (J,T)\exp (-E_{0}/T)\ln (L_{\phi
}/l),  \label{a2}
\end{equation}%
where $l\sim (J\min (J,T))^{\frac{1}{2}}\tau $ is the mean free path
(measured in lattice periods), and $L_{\phi }$ is the Cooper-pair
dephasing length determined by their recombination, interaction with
phonons, emission of Cooper-pair dipoles, etc. In the present paper
we assume that the temperature is low enough ($T\ll E_d$) and thus
$L_{\phi }\gg l$. At $T\ll J$ the pre-exponential factor in
Eq.~(\ref{a2}) equals $-4e^{2}/\pi ^{2}$. At low temperatures or
strong
disorder the correction $\delta \sigma _{WL}$ becomes comparable with $%
\sigma $, which corresponds to the strong Anderson localization of the
bosons.

Eqs.~(\ref{a1}) and (\ref{a2}) are the main results of our paper. The
conductivity of a JJA in the insulating phase is similar to that for
electrons in disordered metals but contains in addition the activation
exponent determining the density of the excited Cooper pairs.

Now we formulate the model and derive the above results. We start with a
standard effective Hamiltonian $\hat{\mathcal{H}}$~\cite{Efetov:fundamental}
describing the motion of Cooper pairs in a JJA
\begin{equation}
\hat{\mathcal{H}}=\frac{1}{2}\sum_{i,j}B_{ij}\hat{n}_{i}\hat{n}%
_{j}-\sum_{i,j}J_{ij}\cos (\phi _{i}-\phi _{j}),  \label{Hamiltonian}
\end{equation}%
where the indices $i$ and $j$ label the superconducting islands,
$B_{ij}$ is the inverse capacitance matrix of the array in units of
$(2e)^{2}$, $\phi _{i}$ and $\hat{n}_{i}=-i\partial /\partial
\phi_{i}$ are respectively the phase of the superconducting order
parameter and the operator of the number of excess Cooper pairs in
the island $i$. $J_{ij}$ is the energy
of Josephson coupling between neighboring islands $i$ and $j$, $%
J_{ij}=J_{ji}=J$. In these notations, adding a Cooper pair to the
site $i$ requires the charging energy $B_{ii}/2$, which is
site-dependent.
 The
average value of this energy is $B/2\equiv \langle B_{ii}/2\rangle
$. We consider the JJA deeply in the insulating phase, $J\ll B$.

Irregularities of the array can be described by the fluctuations
$\delta B_{ij}$ and $\delta J_{ij}$. As we show below, current in
the array is carried by the individual bosons or antibosons
describing respectively excess Cooper pairs on the islands or
\textquotedblleft Cooper-pair-holes\textquotedblright. Random offset
charges weakly coupled to the array may shift the energy of a Cooper
pair on an island, and, thus, contribute to the fluctuations of the
coefficients $B_{ii}$.

Neglecting the Josephson couplings and disorder one obtains a discrete
spectrum of the excitation energies of the system determined by the eigenvalues
of the first term in Eq.~(\ref{Hamiltonian}). The eigenvalues of the
operators $\hat{n}_{i}$ are integers. The ground state corresponds to all $%
n_{i}=0$. All excited states are degenerate as long as the system remains
translationally invariant. Of course, no \textit{dc} current can flow
through the system in this limit.

The degeneracy of the excited states is lifted in the presence of the
Josephson tunnelling $J_{ij}$ between the islands. The Hamiltonian, Eq. (\ref%
{Hamiltonian}), is equivalent to a tight-binding model for bosons: their
states form a band with a width proportional to $J.$ As a result,
macroscopic \textit{dc }transport is possible. The disorder results in the
scattering of the bosons inside the bands and leads to the finite
conductivity, Eqs.~(\ref{a1}) and (\ref{a2}).

At low temperatures the \textit{dc }current is carried by bosons with the
charge $2e$ and antibosons with the charge $-2e.$ The conductivity is
dominated by the lowest energy bands of bosonic and antibosonic states.
These two bands are located near the energy $B/2$ \cite{remark}.

In order to calculate the conductivity at low temperatures we may thus
consider a reduced Hilbert space: the $i$-th island has only three quantum
states: $|0\rangle _{i}$, zero excess Cooper pairs on it, and $|\pm 1\rangle
_{i}$, one Cooper pair (antipair). It is convenient to rewrite the
Hamiltonian (\ref{Hamiltonian}) in this space in terms of pseudospin
operators $\hat{S}_{i}^{+},$ $\hat{S}_{i}^{-}$ and $\hat{S}_{i}^{z}$: $\hat{S%
}_{i}^{\pm }|0\rangle _{i}=\sqrt{2}|\pm 1\rangle _{i}$, $\hat{S}_{i}^{\pm
}|\mp 1\rangle _{i}=\sqrt{2}|0\rangle _{i}$, $\hat{S}_{i}^{\pm }|\pm
1\rangle _{i}=0,$ $\hat{S}_{i}^{z}|0\rangle _{i}=0,$ $\hat{S}_{i}^{z}|\pm
1\rangle _{i}=\pm |\pm 1\rangle _{i}$ corresponding to the pseudospin $%
S_{i}=1.$ The reduced Hamiltonian
\begin{equation}
\hat{\mathcal{H}}_{red}=\frac{1}{2}\sum_{i,j}B_{ij}\hat{S}_{i}^{z}\hat{S}%
_{j}^{z}-\frac{1}{2}\sum_{i,j}J_{ij}\hat{S}_{i}^{+}\hat{S}_{j}^{-}
\label{redham}
\end{equation}%
is equivalent to an anisotropic Heisenberg spin-$1$ model. The pseudospin
operators obey the conventional commutation relations%
\begin{equation}
\left[ \hat{S}_{i}^{+},\hat{S}_{j}^{-}\right] =2\delta _{ij}\hat{S}_{i}^{z},%
\text{\quad }\left[ \hat{S}_{i}^{z},\hat{S}_{j}^{\pm }\right] =\pm \delta
_{ij}\hat{S}_{i}^{\pm }.  \label{commutation}
\end{equation}

We assume for simplicity that the islands in the array form a square
lattice and begin with {calculating} the excitation spectrum in an
ideal JJA without disorder. The states of the Hamiltonian
$\hat{\mathcal{H}}_{red}$ can be classified by the
$S^{z}$-projection of the total spin. The ground state corresponds
to $S_{i}^{z}=0$ for all $i.$

In order to calculate the conductivity we consider states corresponding to a
single boson or antiboson in the array ($S^{z}=\pm 1$). For these states the
eigenenergy of the first term of the Hamiltonian $\hat{\mathcal{H}}_{red}$,
Eq. (\ref{redham}), equals $B/2.$ In the limit $J\ll B$, we approximate~the
eigenfunction of $\hat{\mathcal{H}}_{red}$ for $S^{z}=\pm 1$ by a plane wave
\begin{equation}
|\mathbf{k}\rangle =N^{-\frac{1}{2}}\sum_{\mathbf{r}}e^{i\mathbf{k}\mathbf{r}%
}|\mathbf{r}\rangle ,  \label{excitationwavefunction}
\end{equation}%
where $N$ is the number of the islands in the array. The corresponding
excitation spectrum takes the form
\begin{equation}
E(\mathbf{k})=B/2-2J\cos k_{x}-2J\cos k_{y},  \label{spectrum}
\end{equation}%
where $\bk=(k_x,k_y)$. Thus, the excitation spectrum has a narrow
band of the width $8J$ separated from the ground state by the gap
\begin{equation}
E_{0}=B/2-4J.  \label{gap}
\end{equation}%

As long as the gap significantly exceeds the temperature the density of the bosons is exponentially small.
In this limit the interaction between them
can be neglected and we can describe the system in terms of a single-particle
 tight-binding Hamiltonian with the spectrum given by Eq. (\ref%
{spectrum}). Disoder manifests itself in the model through the
fluctuating parts $\delta B$ and $\delta J$ of the parameters $B$ and $J$.

The conductance of the array can be calculated using the
standard Kubo linear-response theory. The operator $\hat{I}_{ij}$ of the
current between the $i$-th and $j$-th islands reads
\begin{equation}
\hat{I}_{ij}=ieJ_{ij}\left(
\hat{S}_{j}^{+}\hat{S}_{i}^{-}-\hat{S}_{i}^{+}\hat{S}_{j}^{-}\right),
\label{current}
\end{equation}%
Its expectation value $I_{ij}\left( \omega \right) $ can be expressed through%
\textbf{\ }{the retarded correlation function} of currents $\Pi _{ij,kl}(\omega)$:
\begin{equation}
\Pi _{ij,kl}(\omega )=\frac{1}{2}\left. \int_{-\beta }^{\beta }\langle \hat{I%
}_{ij}(\tau )\hat{I}_{kl}(0)\rangle e^{i\Omega _{n}\tau }d\tau
\right\vert _{i\Omega _{n}\rightarrow \omega +i0},  \label{a6}
\end{equation}%
\begin{equation}
I_{ij}(\omega )=-i{\omega }^{-1}\sum_{(kl)}\left( \Pi
_{ij,kl}(\omega )-\Pi _{ij,kl}(0)\right)
(\mathbf{E}\mathbf{l}_{kl}), \label{bondconductance}
\end{equation}
{where {\bf E} is the electric field, $\bl_{kl}$-- vector connecting
islands $k$ and $l$.}

Depending on whether the length $L_x$ of the array
 is smaller or larger than the mean free path $l$, the transport in the sample is
respectively ballistic or diffusive.

Using Eqs.~(\ref{a6}) and (\ref{bondconductance}) in the ballistic
limit we find the conductance of an $L_{x}\times L_{y}$ rectangular
array
\begin{equation}
G=\frac{8e^2}{\pi}L_y\sinh \left( {2J}/{T}\right) I_{0}\left(
{2J}/{T}\right) \exp \left( -E_{0}/T\right).
\label{arrayconductance}
\end{equation}
Here $I_{0}$ is the modified Bessel function.
Of course, at $l\sim L_x$ Eq.~(\ref{arrayconductance}) matches the diffusive conductance
$\sigma L_y/L_x$ with $\sigma$ given by Eq.~(\ref{a1}).

Let us note now that in the low-temperature limit the array
resembles a conventional semiconductor. Indeed, at low density of
the bosons the particle statistics is not important. The bosons thus
can be considered as doubly charged electrons thermally activated to
the conduction band of the width $8J$, the latter being separated
from the valence band of the semiconductor by the gap $E_{0}$,
Eq.~(\ref{gap}). Hence, one can evaluate the conductivity in the
diffusive regime using the standard diagrammatic
technique~\cite{AGD}. We assume that the disorder is weak enough for
the elastic scattering time $\tau$ to exceed the inverse
characteristic kinetic energy $\min (J,T)$ of the bosons in the
conduction band:
\begin{equation}
\tau (\min (J,T))\gg 1.  \label{a0}
\end{equation}%

{ To average} over the disorder {we assume for simplicity} a
Gaussian distribution
for deviations $\delta B_{ii}$ and $\delta J_{ij}$ from the average values $%
B $ and $J$ with correlations
\begin{equation*}
\langle \delta B_{ii}\delta B_{jj}\rangle =f_{1}\delta
_{ij},\quad\langle \delta J_{ij}\delta J_{kl}\rangle =f_{2}(\delta
_{ik}\delta _{jl}+\delta _{il}\delta _{jk}).
\end{equation*}
{Our results Eqs. (\ref{a1}) and (\ref{a2}) with appropriate $\tau$
apply nevertheless for arbitrary not short-correlated distributions
of the fluctuations.}

The basic element of the perturbation theory is the contraction rule for the
effective disorder potential $\hat{V}$
\begin{eqnarray}
\langle \hat{V}_{\mathbf{k}_{1}\mathbf{p}_{1}}\hat{V}_{\mathbf{k}_{2}\mathbf{%
p}_{2}}\rangle &=&(2\pi )^{2}\gamma (\mathbf{k}_{1}+\mathbf{k}_{2},\mathbf{k}%
_{1}-\mathbf{p}_{2})  \notag \\
&&\delta (\mathbf{k}_{1}+\mathbf{k}_{2}-\mathbf{p}_{1}-\mathbf{p}_{2}),
\label{a11} \\
\gamma (\mathbf{k}_{+},\mathbf{k}_{-}) &=& f_{1}/4+f_{2}\sum_{i}%
\left( e^{\mathbf{l}_{i}\mathbf{k}_{+}}+e^{\mathbf{l}_{i}\mathbf{k}%
_{-}}\right) ,  \notag
\end{eqnarray}%
where $\mathbf{l}_{i}$ ($i=1\ldots 4$) are the unit vectors connecting an
island with its nearest neighbors.

The standard procedure (see e.g. Ref.~\cite{AGD}) of evaluation of $\tau$ in the limit $k\ll 1$
under the condition (\ref{a0}) gives
\begin{equation}
\tau ^{-1}=\left( f_{1}/8+4f_{2}\right) /J.  \label{a13}
\end{equation}%
The scattering time $\tau $
remains of the same order of magnitude at arbitrary momentum $k\sim1$.

We use the Kubo-Greenwood formula
\begin{eqnarray}
\sigma _{\alpha \beta } &=&\frac{2(2e)^{2}}{\omega }\int \frac{d\mathbf{p}}{%
(2\pi )^{2}}\int \frac{d\varepsilon }{2\pi }\left( n(\varepsilon
)-n(\varepsilon +\omega )\right)  \notag \\
&&\left\langle v_{\alpha }G^{A}(\mathbf{p},\varepsilon )v_{\beta }G^{R}(%
\mathbf{p},\varepsilon )\right\rangle ,  \label{a14}
\end{eqnarray}%
where $\mathbf{v}=2J(\sin k_{x},\sin k_{y})$ is the velocity of the boson, $%
n(\varepsilon )\approx \exp (-\varepsilon /T)$ is the Boltzman
distribution of the excitations in the conduction band and $
G^{R,A}(\mathbf{k},\varepsilon)$ are retarded (advanced) Green's
functions of non-interacting bosons, $\langle
G^{R,A}(\mathbf{k},\varepsilon)\rangle=(\varepsilon
-E(\mathbf{k})\pm i/(2\tau ))^{-1}$. Using the condition (\ref{a0})
we come to Eq.~(\ref{a1}).

We emphasize that Eq.~(\ref{a1}) has been obtained in the standard scheme
neglecting diagrams with crossing impurity lines \cite{AGD}. In disordered
metals this approach is applicable in the limit $\varepsilon _{F}\tau \gg 1$,
where $\varepsilon _{F}$ is the Fermi energy.
Here the role of large parameter is played by $\tau (\min (J,T))$ [cf. Eq.~(\ref{a0})].

Next we calculate the weak localization correction $\delta\sigma
_{WL}$ to the conductivity. Again, the condition (\ref{a0}) allows
us to repeat the
summation of the diagrams of Ref.~[\onlinecite{GLK}] and to arrive at Eq.~(%
\ref{a2}).

The limiting case $C/C_{0}\gg 1$, where $C$ is the mutual
capacitance of neighboring islands and $C_{0}$ is the self-capacitance, often
corresponds to the experimental situation (see, e.g., Ref.~\cite{Mooji}) and
is especially interesting from the theoretical point of view.
In this case the charging energy $E_0$ of a single boson significantly exceeds the dipole energy $E_d$,
$E_0/E_d\sim\ln(C/C_0)\gg1$.

Such bosons resemble vortices in superconductors \cite{De Gennes}. A finite
ratio $C/C_{0}$ determines the scale of the interaction of the bosons and
plays the same role as the penetration depth cutting the logarithmic
interaction of vortices in superconductors. In the limit $%
C/C_{0}\rightarrow \infty $ the energy of the bosons logarithmically
diverges with the size $L$ of the sample. Using the analogy with the
vortices one can expect in this limit the BKT transition,
Ref.~[\onlinecite{BKT}], with the critical temperature $T_{K}$ of
order of the dipole energy $E_{d}$.

Properties of the system of the bosons are similar to those of the
system of vortices in conventional $2D$ superconductors. In the BKT
scenario single vortices do not enter the system below $T_{K}$ and
the system is a superfluid. In the JJA considered here, there are no
single bosons in the limit $C/C_{0}\rightarrow \infty $ and the
conductivity vanishes. However, the BKT transition in
superconductors is known to smear because the energy of a single
vortex is finite due to the finite penetration depth, which makes
vortices itinerant resulting in a finite resistivity. In the JJA the
energy of the bosons is finite due to the finite $C/C_{0}$ and their
motion makes the electric transport possible. As the density of the
bosons is proportional to $\exp(-E_0/T)$, so is the conductivity,
Eq.~(\ref{a1}). At temperatures $T\ll E_d\sim T_K$ one can neglect
the presence of dipole excitations in the array which could lead to
an additional scattering or relaxation of the single bosons and to a
screening of Coulomb interaction.

At $T>T_{K}$ single vortices in 2D superconductors exist due to the
entropy contribution, Ref.~[\onlinecite{BKT}], and the superfluidity
is destroyed. 2D superconductors, strictly speaking, do not exist.
Thus, as the superfluidity below $T_{K}$ is either absent, one
obtains a crossover from the finite exponentially small resistivity
contributed by the motion of single vortices to the resistivity of
normal metals. Analogously, in a JJA one can expect a crossover from
the exponentially low conductivity at $T\ll E_{d}$, Eq. (\ref{a1}),
to a temperature-independent conductivity at $T>E_{d}$. In other
words, the Coulomb blockade effects are important at low temperature
$T<E_{d}$ but can be neglected at $T>E_{d}$. Interference of the
bosons results in the existence of one more temperature region where
effects of localization can play an important role and the Coulomb
blockade is further enforced by the Anderson localization.

Recently, a similar model of a $2D$ JJA (but without disorder) has
been suggested in Refs.~[\onlinecite{Fistul:superinsulator}] and
[\onlinecite{Vinokur:superinsulator}] to describe the experiments on
strongly disordered superconductors~\cite{Shahar,
Vinokur:superinsulator}. The authors obtained an exponential
behavior of the conductivity with the activation gap $B/2$ for
$T>E_{d}$ and a double exponential behavior for $T\ll E_{d}$.
Clearly, our results, Eqs.~(\ref{a1}), (\ref{a2}), and the absence
of the Coulomb blockade at $T>E_{d}$ are in a strong disagreement
with those findings. A detailed criticism of the theoretical
treatment of Refs.~[\onlinecite{Fistul:superinsulator}] and
[\onlinecite{Vinokur:superinsulator}] can be found in
Ref.~[\onlinecite{Efetov:comment}].
A comparison of our results with the results of{\ relevant
}experimental works will be presented elsewhere. Also, it would be
interesting to clarify the question of the many-body localization
\cite{Basko} of the bosons in the system under consideration.

In conclusion, we calculated the \textit{dc} conductivity of a large
Coulomb-blockaded Josephson junction array at low temperatures and
demonstrated that it is determined by the thermally activated
single-Cooper-pair excitations on the superconducting islands. In
the absence of macroscopic structural disorder in the array the
transport is ballistic. In the presence of sufficiently weak
disorder the conductivity in a certain temperature range is
described by the Drude-type formula multiplied by the activation
exponent $\exp (-E_{0}/T)$, where $E_{0}$ is the Coulomb energy of
the single-Cooper-pair excitation. Lowering the temperature below
some characteristic value results in even faster decrease of the
conductivity due to the Anderson localization of Cooper pairs.

The authors thank M.~Yu. Kharitonov for discussions. The work has
been financially supported by SFB Transregio 12, SFB 491, and US DOE
contract No. DE-AC02-06CH11357.


\end{document}